\documentclass[journal,final,twocolumn]{IEEEtran}
\usepackage{graphicx}
\usepackage{bm}
\usepackage{url}
\usepackage{amsfonts,amssymb,amsmath} 
\usepackage{flushend}
\usepackage{setspace}
\addtocounter{MaxMatrixCols}{50}
\allowdisplaybreaks

\begin{document}

\title{Rate-Compatible Protograph-based LDPC Codes for Inter-Symbol
  Interference Channels}

\author{Thuy Van Nguyen, {\em Student Member, IEEE,} Aria Nosratinia,
  {\em Fellow, IEEE,} and Dariush Divsalar, {\em Fellow,
    IEEE}\thanks{
    Thuy Van Nguyen and Aria Nosratinia are with the Department of
    Electrical Engineering, The University of Texas at Dallas,
    Richardson, TX 75083-0688 USA, Email nvanthuy@utdallas.edu,
    aria@utdallas.edu. Dariush Divsalar is with the Jet Propulsion
    Laboratory, California Institute of Technology, Pasadena, CA
    91109-8099 USA Email: Dariush.Divsalar@jpl.nasa.gov.}}

\maketitle

\begin{abstract} 
This letter produces a family of rate-compatible protograph-based LDPC
codes approaching the independent and uniformly distributed (i.u.d.)
capacity of inter-symbol interference (ISI) channels. This problem is
highly nontrivial due to the joint design of structured
(protograph-based) LDPC codes and the state structure of ISI
channels. We describe a method to design nested high-rate protograph
codes by adding variable nodes to the protograph of a lower rate
code. We then design a family of rate-compatible protograph codes
using the extension method. The resulting protograph codes have
iterative decoding thresholds close to the i.u.d.\ capacity. Our
results are supported by numerical simulations.

\end{abstract}

\section{Introduction}

Rate-compatibility is a desirable feature that allows a single
encoder/decoder operating over a variety of code rates, which can be
selected to match channel conditions.  While rate-compatible LDPC
codes have been studied for memoryless AWGN channels, in
channels with memory, especially inter-symbol-interference (ISI)
channels, the problem of rate-compatible LDPC coding has been open
until now.

The capacity of ISI channels with finite-alphabet inputs remains
open. If the source is restricted to be independent and uniformly
distributed (i.u.d.), the capacity of binary-input ISI channels, also
called partial response channels, is known as the
i.u.d.\ capacity~\cite{Kavcic:IT03} (also known as symmetric
information rate). Irregular LDPC codes have been designed to approach
the i.u.d.\ capacity of partial response
channels~\cite{McLaughlin:TMAG02,Varnica:COML03,Pfister:IT07,Franceschini:EXIT05},
but they generally lack a structure to enable easy encoding and fast
decoding. In this letter, we address the problem of designing
rate-compatible structured LDPC codes that are capacity-approaching
for ISI channels.

Our structured LDPC codes are built from a small graph, called a
protograph~\cite{Thorpe2003}. Protograph codes have demonstrated very
good performance in terms of iterative decoding thresholds as well as
finite-length performance with low encoding and decoding
complexity~\cite{Thorpe2004,Divsalar2009} in the AWGN
channels. Protograph codes have been designed for several partial
response channels~\cite{Thuy:ICC12,Fang:TCOM12}. The authors of the
present paper in~\cite{Thuy:ICC12} proposed a method to design a rate-$1/2$
capacity-approaching protograph code in ISI
channels. Fang et al.~\cite{Fang:TCOM12} designed a
nested protograph family that is good for the dicode and EPR4 channels
based on finite-length EXIT analysis.  However, the prior activity in the
open literature did not produce {\em rate-compatible} LDPC codes for
the ISI channel.

The main contribution of this paper is a set of rate-compatible structured
codes based on protographs that are i.u.d.\ capacity
approaching. We design two families of protograph codes: the nested
high-rate protographs where high-rate codes are built from a low-rate
protograph by adding more variable nodes; and the rate-compatible
protographs where information payload of all members are
identical. All proposed codes have thresholds that are within
a gap of $0.5$ dB to i.u.d.\ capacity. The performance of the
proposed codes with $16$k data over the dicode and EPR4 channels is
reported. The performance of our codes exhibits a frame error rate of
$3\times 10^{-6}$ at a gap of $1.1$ dB from the i.u.d.\ capacity limits.

\section{Protograph Design for ISI Channels}
\label{sec:design}

A protograph~\cite{Thorpe2003} is a Tanner graph with a relatively
small number of nodes. A protograph code (an equivalent LDPC code) is
a larger derived graph constructed by applying a
``copy-and-permutation'' operation on a protograph, a process known as
\emph{lifting}. In this lifting, the protograph is copied $N$ times,
then a large LDPC code graph is obtained by permuting $N$
variable-to-check pairs (edges), corresponding to each of the edge types
of the original protograph.

Designing good protographs for the AWGN channel has been addressed
in~\cite{Divsalar2009, Thuy:Tcom11}. Its extension to ISI channels is
not easy due to its concatenation with a BCJR equalizer. We proposed
in~\cite{Thuy:ICC12} a direct method to design a rate-$1/2$
protograph in partial response channels. However, practical systems,
e.g., in magnetic recoding, require LDPC codes with rates up to
$0.9$. Applying the method of~\cite{Thuy:ICC12} to design high-rate
codes is impractical due to the high dimensionality of
protographs. Another approach to avoid this problem is to exploit a
nested structure which allows building a high-rate code from a low-rate
code.  In this letter, we consider the same system model and receiver
structure as in~\cite{Thuy:ICC12} where two popular
channels, the dicode channel $h(D) = (1-D)/\sqrt{2}$ and the EPR4
channel $h(D) = (1+D-D^2-D^3)/2$, are studied.

Let us first focus on designing a rate-$1/2$ code. We do not use the
rate-$1/2$ protograph reported in~\cite{Thuy:ICC12} because it is too
large\footnote{Large protographs may require many intermediate steps
  to get a rate of $0.9$.} for our goal of building nested codes with
rates up to $0.9$. We apply the method in~\cite{Thuy:ICC12} to design
a smaller rate-$1/2$ graph, thus making it easier to design good
high-rate, rate-compatible codes in the next section.

\begin{table}
\caption{Threshold ($E_b/N_0$ dB) of new codes}
\centering
\small \begin{tabular}{|c|c|c|c|c|}
\hline
 &\multicolumn{2}{c|}{dicode}&\multicolumn{2}{c|}{EPR4}\\
\hline
Code & Code& Gap to &  Code& Gap to\\
Rate & thres. &cap.& thres. & cap.\\
 \hline
\multicolumn{5}{|c|}{The nested family}\\ \hline
1/2 & 1.3&0.5&1.7&0.5\\ \hline
2/3 & 2.2& 0.4&2.6&0.4\\ \hline
3/4 & 2.7& 0.3&3.2&0.4\\ \hline
4/5 & 3.1& 0.3&3.6&0.4\\ \hline
5/6 & 3.3& 0.2&3.9&0.3\\ \hline
6/7 & 3.6& 0.2&4.1&0.3\\  \hline 
7/8 & 3.8& 0.2&4.3&0.3\\ \hline
8/9 & 4.0& 0.3&4.2&0.3\\ \hline
9/10 &4.2& 0.3&4.7&0.4\\ \hline
\multicolumn{5}{|c|}{The rate-compatible family}\\ \hline
9/10  & 4.2& 0.3&4.7&0.4\\ \hline
27/31 & 4.0 & 0.4&4.3&0.3\\ \hline
27/32 & 3.6 & 0.3&4.0&0.3\\ \hline
27/33 & 3.3 & 0.3&3.7&0.3\\ \hline
27/34 & 2.9 & 0.2&3.5&0.3\\ \hline
27/35 & 2.9 & 0.3&3.3&0.3\\ \hline
27/37 & 2.6 & 0.3&3.0&0.3\\  \hline 
27/39 & 2.3 & 0.3&2.7&0.3\\ \hline
27/41 & 2.1 & 0.3&2.5&0.3\\ \hline
\end{tabular}
\label{ta:threshold}
\end{table}

To demonstrate, let us search for a rate-$1/2$ protograph that
contains $3$ check nodes and $6$ variable nodes without any punctured
node, as pointed out in~\cite{Thuy:ICC12}. A good protograph should
contain nodes of both degree 1 and 2. Thus, to greatly reduce the
search space, we start by a search structure with one degree-1
and one degree-2 variable node in the form of the protomatrix
\begin{equation}
H_{search}^{1/2}=
\begin{pmatrix}
1&0&x_1& x_4& x_7  & y_{1} \\
0&1&x_2& x_5& x_{8}& y_{2} \\
0&1&x_3& x_6& x_{9}& y_{3}
\end{pmatrix} \; ,
\end{equation}
where $x_i$, $i=1,\ldots,9$ and $y_j$, $j=1,\ldots,3$ , are the number
of edges connecting their associated row (check node) and column
(variable node) in which $y_1, y_2, y_3$ correspond to the highest degree
variable node. In order for the code to have the linear minimum
distance growth property, the edge summation over the last two rows
within the last $4$ columns should be $3$ or higher~\cite{Divsalar_ITA10},
i.e., $x_2+x_3\geq 3$, $x_5+x_6\geq 3$, $x_{8}+x_{9}\geq 3$ and
$y_{2}+y_{3}\geq 3$. We can further simplify the problem by limiting
$x_i\in\{0,1,2\}$ and $y_j\in\{1,2,3,4\}$. Our objective in this
specific example is to find a protograph that has the lowest iterative
decoding threshold over the dicode channel.

\begin{figure}
\centering
\includegraphics[width=2.5in]{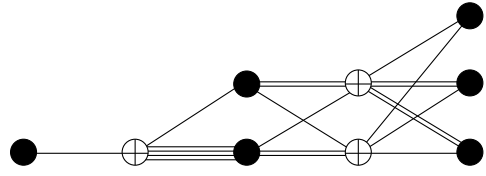}
\caption{The rate-$1/2$ protograph optimized for the dicode channel}
\label{fig:isi}
\end{figure}

 After a simple search, the resulting protograph, called the ISI code,
 is in the form of
\begin{equation}
H^{1/2}_{ISI}=
\begin{pmatrix}
1& 0& 0& 1& 0& 4 \\
0& 1& 2& 1& 2& 2 \\
0& 1& 1& 2& 1& 1
\end{pmatrix} \; .
\end{equation}
The search is performed using PEXIT (protograph EXIT)
analysis~\cite{Liva2007}. The designed protograph is shown in
Fig.~\ref{fig:isi}, and the corresponding thresholds in a variety of
channels are given in Table~\ref{ta:threshold}. This code has a
threshold within $0.5$ dB of the i.u.d.\ capacity limit of the dicode
channel.

\section{Rate-Compatible Codes for ISI Channels}
\label{sec:RC_codes}

There are two popular methods of building rate-compatible codes. One
of them is via puncturing, i.e., starting from a low-rate code and
selectively deleting parity bits to build high-rate codes. The other
method involves extension, i.e., starting from a high-rate code,
low-rate codes are built by adding the same number of variable and
check nodes. The former method in general is not good for channels
with memory, as pointed out by~\cite{Thuy:ICC12} except when only
degree-$1$ variable nodes are punctured~\cite{Fang:TCOM12}. The latter
method will be applied to our rate-compatible code design in this
section.

\subsection{Nested High-rate Codes}
In order to use the extension method to design rate-compatible codes,
we first need to design a high-rate code. We use a nested lengthening
structure as follows
\begin{equation}
H_{h} = [H_{l} \;\; H_e] \; ,
\end{equation}
where $H_{h}$ and $H_{l}$ are protomatrices of a
high-rate and low-rate protograph, respectively, $H_e$ is an extension
matrix whose columns are the newly added variable nodes and its elements are
the number of edges connecting a new variable node to an existing check node.

In this example, we start from the rate-$1/2$ protograph designed in
Section~\ref{sec:design}. We then design a family
of nested codes with rates in the form of $R=\frac{n}{n+1}$,
$n=1,2,\ldots$. Within this nested family, successively higher rate
protographs are constructed by adding 3 new variable nodes with each
step.\footnote{If we use the rate-$1/2$ code in~\cite{Thuy:ICC12}, $4$
  new variable nodes are needed instead.} We then apply the same
search method used in Section~\ref{sec:design} to find the protographs
with the lowest threshold. We then are able to design the code with
rate up to $9/10$. Due to space limitation, we only present the result
of the rate $0.9$ code in which the other eight codes are deduced. The
rate-$0.9$ code protomatrix is 
\setlength{\arraycolsep}{0.045cm}
\begin{align}
&H_{0.9}=\nonumber\\
&\left(
\begin{array}{cccccc|ccc|ccc|ccc|ccc|ccc|ccc|ccc|ccc}
1&0&0&1&0&4&2&0&0&0&0&2&2&0&0&0&0&1&0&0&0&2&0&0&0&0&0&0&0&1\\ 
0&1&2&1&2&2&1&2&2&2&1&2&2&2&2&2&2&1&2&2&2&1&2&1&1&1&1&2&2&1\\ 
0&1&1&2&1&1&2&1&1&1&2&1&1&1&1&1&1&2&2&1&1&2&1&2&2&2&2&2&1&2 
\end{array}
\right)
\label{eq:R910}
\end{align}
In the above protomatrix, we separate each code rate level by a line
as shown in~\eqref{eq:R910}. The thresholds of these nine
nested codes are shown in Table~\ref{ta:threshold} where the high-rate
codes have their iterative decoding thresholds within $0.4$ dB of the i.u.d.
capacity of dicode and EPR4 channels. Considering this small gap to i.u.d.
capacity of ISI channels, we can say that nested high-rate codes are
good enough and there is no need to design a high-rate code directly as
described in Section~\ref{sec:design}.

\subsection{Rate-Compatibility via Extension}

Using the extension method, the low rate protomatrix is in the
following form

\begin{equation}
H_{l} = \left(
\begin{array}{cc}
H_1\quad 0\\
A\quad B
\end{array}
\right) \; ,
\end{equation}
where $H_l$ and $H_1$ are the low-rate and high-rate protomatrices,
respectively, $0$ is the zero matrix, $A$ is the matrix whose elements
are number of edges connecting new check nodes to the existing variable
nodes, and $B$ is the matrix whose elements are the number of edges
connecting between new check nodes to new variable nodes. To simplify
the problem, we assume that $B$ is identity.

Starting with the rate-$9/10$ code, we design a family of
rate-compatible codes as follows. Each time, we add one variable node
and one check node to the protomatrix of the high-rate code. The new
codes have the rates in the form of $R=\frac{27}{30+m}$, where $m$ is
the number of variable and check nodes added into the rate-$9/10$
code. Due to space limitation, we only design nine rate-compatible
codes with rates from $27/41$ to $9/10$, where the biggest graph/code
has the lowest rate of $R=27/41$. Equation~\eqref{eq:RC-codes} shows
its protomatrix which contains the other eight rate-compatible code
protomatrices. Iterative decoding thresholds of these rate-compatible
codes over the dicode and EPR4 channels are shown in
Table~\ref{ta:threshold}. Again, all these codes can operate closely
to capacity with thresholds gaps of $0.4$ dB to i.u.d.\ capacity
limits.

\begin{figure*}
\begin{spacing}{1.1}
\setlength{\arraycolsep}{0.07cm}
\begin{align}
H_{14\times 41} =
\begin{pmatrix}
1&0&0&1&0&4&2&0&0&0&0&2&2&0&0&0&0&1&0&0&0&2&0&0&0&0&0&0&0&1&0&0&0&0&0&0&0&0&0&0&0\\ 
0&1&2&1&2&2&1&2&2&2&1&2&2&2&2&2&2&1&2&2&2&1&2&1&1&1&1&2&2&1&0&0&0&0&0&0&0&0&0&0&0\\ 
0&1&1&2&1&1&2&1&1&1&2&1&1&1&1&1&1&2&2&1&1&2&1&2&2&2&2&2&1&2&0&0&0&0&0&0&0&0&0&0&0\\ 
0&0&1&1&1&1&1&0&1&0&0&1&0&0&0&0&0&0&0&0&0&0&0&1&1&0&0&1&0&0&1&0&0&0&0&0&0&0&0&0&0\\ 
0&0&1&1&1&2&0&0&1&0&1&0&1&1&1&0&1&0&0&0&0&0&0&0&0&0&0&0&0&1&0&1&0&0&0&0&0&0&0&0&0\\ 
0&0&1&1&1&1&1&0&1&1&0&1&1&0&1&1&1&0&0&0&0&0&0&0&0&0&0&0&0&0&0&0&1&0&0&0&0&0&0&0&0\\ 
0&1&0&0&0&2&0&0&0&0&1&1&1&0&1&0&1&0&0&0&0&1&0&1&0&0&0&1&0&1&0&0&0&1&0&0&0&0&0&0&0\\ 
0&0&1&0&1&1&0&0&0&0&0&0&1&1&0&0&0&0&0&0&0&0&1&0&0&1&0&0&0&1&0&0&0&0&1&0&0&0&0&0&0\\ 
0&0&0&0&0&2&0&0&1&0&0&0&1&0&0&0&1&0&0&0&0&1&1&0&1&0&0&0&0&1&0&0&0&0&0&1&0&0&0&0&0 \\
0&0&0&0&0&2&0&0&0&0&0&1&0&1&0&0&0&0&0&0&1&0&0&0&0&0&0&1&0&1&0&0&0&0&0&0&1&0&0&0&0\\ 
0&0&0&0&0&1&0&0&0&0&0&1&1&1&0&0&0&0&0&0&0&1&0&0&0&0&0&1&1&1&0&0&0&0&0&0&0&1&0&0&0 \\
0&0&0&0&0&2&0&0&1&0&0&1&0&0&0&0&1&0&0&0&1&1&0&0&0&0&1&0&0&1&0&0&0&0&0&0&0&0&1&0&0\\ 
0&0&0&0&0&1&0&0&0&0&0&0&0&0&0&0&1&0&0&0&1&0&0&0&1&0&0&1&1&1&0&0&0&0&0&0&0&0&0&1&0\\ 
0&0&0&0&0&1&1&0&0&1&0&0&0&0&0&1&0&0&0&0&1&0&0&0&1&0&1&1&0&0&0&0&0&0&0&0&0&0&0&0&1
\end{pmatrix}
\label{eq:RC-codes}
\end{align}
\end{spacing}
\end{figure*}

\section{Numerical Results}
\label{sec:numerical}

Our protograph codes are derived from protomatrices (protographs)
designed in Section~\ref{sec:RC_codes} in two lifting steps. First,
the protograph is lifted by a factor of $4$ using the progressive edge
growth (PEG) algorithm~\cite{Hu03_PEG} in order to remove all parallel
edges. Then, a second lifting using the PEG algorithm was performed to
determine a circulant permutation of each edge class that would yield
the desired code block length. A circulant lifting of the protograph
results in an overall code that is quasi-cyclic, which is
known~\cite{SmarandacheVontobel:IT2012} to have an upper bound on
minimum distance that is independent of blocklength. However,
experience shows that even with circulant lifting it is preferable to
use protographs that are designed with linear distance
properties, especially with a two-step lifting where the first step
removes parallel edges.

In this section, protograph codes are simulated with the information
payload of $16$k bits. For nested codes, lifting factors of
codes with rates $1/2$, $2/3$, $3/4$, $4/5$, $5/6$, $6/7$, $7/8$,
$9/10$ are $4\times 1364$, $4\times 683$, $4\times 455$, $4\times
342$, $4\times 273$, $4\times 227$, $4\times 195$, and $4\times 153$,
respectively. For rate-compatible codes, only the lowest rate code
whose protomatrix is shown in~\eqref{eq:RC-codes} is constructed with
the lifting factor of $4\times 153$. Other codes are obtained by
removing coded bits and check equations.

\begin{figure}
\centering
\includegraphics[width=3.75in]{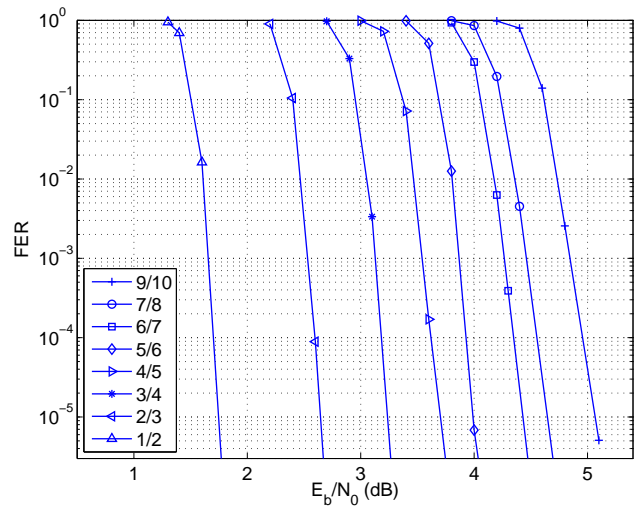}
\caption{Nested protograph family over the dicode channel, payload $16k$}
\label{fig:nested-dicode}
\end{figure}

\begin{figure}
\centering
\includegraphics[width=3.75in]{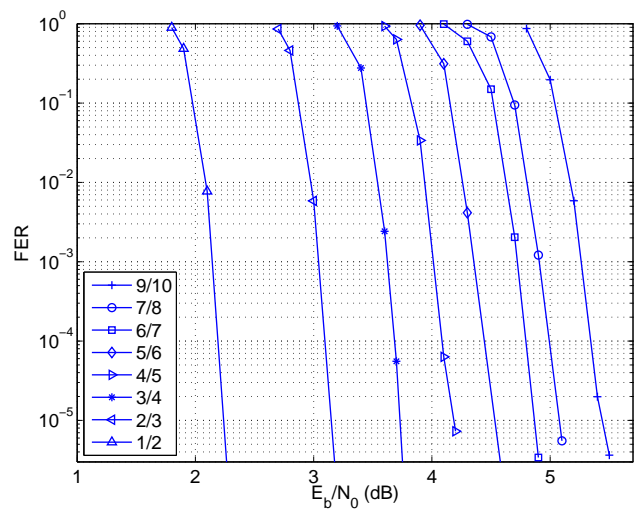}
\caption{Nested protograph family over the EPR4 channel, payload $16k$}
\label{fig:nested-epr4}
\end{figure}

\begin{figure}
\centering
\includegraphics[width=3.75in]{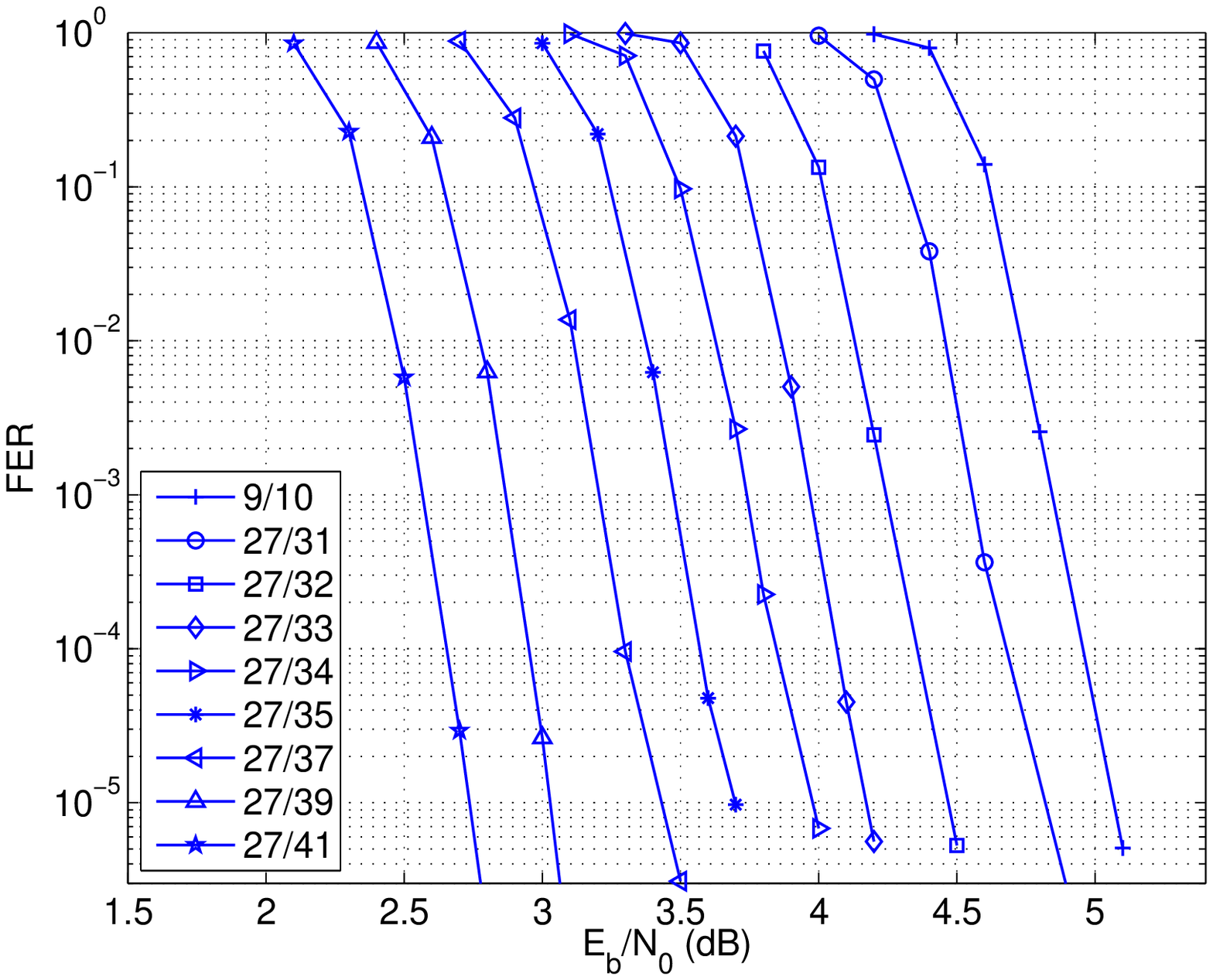}
\caption{Rate-compatible protograph family over the dicode channel}
\label{fig:RC-dicode}
\end{figure}

\begin{figure}
\centering
\includegraphics[width=3.75in]{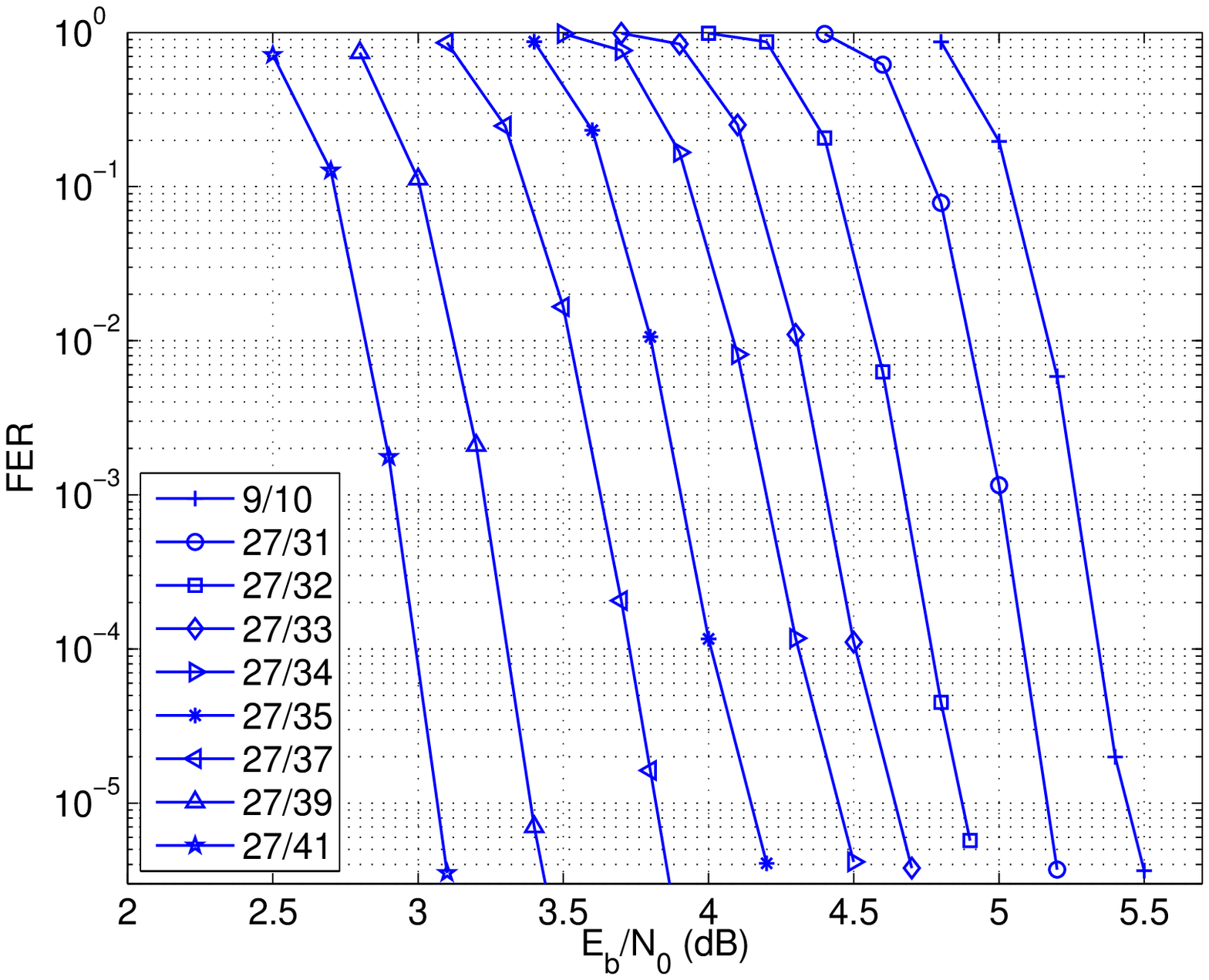}
\caption{Rate-compatible protograph family over the EPR4 channel}
\label{fig:RC-epr4}
\end{figure}

The FER performance of the nested and rate-compatible codes over the
dicode and EPR4 channels are shown in Figs.~\ref{fig:nested-dicode},
\ref{fig:nested-epr4}, \ref{fig:RC-dicode} and~\ref{fig:RC-epr4},
respectively. No error floors are observed down to $FER=3\times 10^{-6}$,
with a gap of $1.1$ dB from the i.u.d.\ capacity.


\section{Conclusion}
\label{sec:conclusion}
This letter presents a design for nested and rate-compatible
protograph-based LDPC codes for ISI channels. Iterative decoding thresholds and
finite length performances of the codes are reported. Analysis and simulation
results show that our codes, which allow easy encoding and fast
decoding, can perform closely to i.u.d.\ capacity limits.

\bibliographystyle{IEEEtran}
\bibliography{IEEEabrv,isi_ref-revised} 

\begin{thebibliography}{10}
\providecommand{\url}[1]{#1}
\csname url@rmstyle\endcsname
\providecommand{\newblock}{\relax}
\providecommand{\bibinfo}[2]{#2}
\providecommand\BIBentrySTDinterwordspacing{\spaceskip=0pt\relax}
\providecommand\BIBentryALTinterwordstretchfactor{4}
\providecommand\BIBentryALTinterwordspacing{\spaceskip=\fontdimen2\font plus
\BIBentryALTinterwordstretchfactor\fontdimen3\font minus
  \fontdimen4\font\relax}
\providecommand\BIBforeignlanguage[2]{{%
\expandafter\ifx\csname l@#1\endcsname\relax
\typeout{** WARNING: IEEEtran.bst: No hyphenation pattern has been}%
\typeout{** loaded for the language `#1'. Using the pattern for}%
\typeout{** the default language instead.}%
\else
\language=\csname l@#1\endcsname
\fi
#2}}

\bibitem{Kavcic:IT03}
A.~Kav$\check{c}$i$\acute{c}$, X.~Ma, and M.~Mitzenmacher, ``Binary intersymbol
  interference channels: {G}allager codes, density evolution, and code
  performance bounds,'' \emph{{IEEE} Trans. Inform. Theory}, vol.~49, no.~7,
  pp. 1636 -- 1652, July 2003.

\bibitem{McLaughlin:TMAG02}
A.~Thangaraj and S.~McLaughlin, ``Thresholds and scheduling for {LDPC}-coded
  partial response channels,'' \emph{{IEEE} Trans. Magn.}, vol.~38, no.~5, pp.
  2307 -- 2309, Sept. 2002.

\bibitem{Varnica:COML03}
N.~Varnica and A.~Kav$\check{c}$i$\acute{c}$, ``Optimized low-density
  parity-check codes for partial response channels,'' \emph{{IEEE} Commun.
  Lett.}, vol.~7, no.~4, pp. 168 -- 170, Apr. 2003.

\bibitem{Pfister:IT07}
J.~B. Soriaga, H.~D. Pfister, and P.~H. Siegel, ``Determining and approaching
  achievable rates of binary intersymbol interference channels using multistage
  decoding,'' \emph{{IEEE} Trans. Inform. Theory}, vol.~53, no.~4, pp. 1416
  --1429, Apr. 2007.

\bibitem{Franceschini:EXIT05}
M.~Franceschini, G.~Ferrari, and R.~Raheli, ``{EXIT} chart-based design of
  {LDPC} codes for inter-symbol interference channels,'' in \emph{Proc. IST
  Mobile and Wireless Communications}, June 2005.

\bibitem{Thorpe2003}
J.~Thorpe, ``Low-density parity-check ({LDPC}) codes constructed from
  protographs,'' IPN Progress Report 42-154, Aug. 2003.

\bibitem{Thorpe2004}
J.~Thorpe, K.~Andrews, and S.~Dolinar, ``Methodologies for designing {LDPC}
  codes using protographs and circulants,'' in \emph{Proc. IEEE ISIT}, July
  2004, pp. 236--236.

\bibitem{Divsalar2009}
D.~Divsalar, S.~Dolinar, C.~R. Jones, and K.~Andrews, ``Capacity-approaching
  protograph codes,'' \emph{{IEEE} J. Select. Areas Commun.}, vol.~27, no.~6,
  pp. 876--888, Aug. 2009.

\bibitem{Thuy:ICC12}
T.~V. Nguyen, A.~Nosratinia, and D.~Divsalar, ``Protograph-based {LDPC} codes
  for partial response channels,'' in \emph{Proc. IEEE ICC}, June 2012, pp.
  2194--2199.

\bibitem{Fang:TCOM12}
Y.~Fang, P.~Chen, L.~Wang, and F.~Lau, ``Design of protograph {LDPC} codes for
  partial response channels,'' \emph{{IEEE} Trans. Commun.}, vol.~60, no.~10,
  pp. 2809--2819, Oct. 2012.

\bibitem{Thuy:Tcom11}
T.~V. Nguyen, A.~Nosratinia, and D.~Divsalar, ``The design of rate-compatible
  protograph {LDPC} codes,'' \emph{{IEEE} Trans. Commun.}, vol.~60, no.~10, pp.
  2841--2850, Oct. 2012.

\bibitem{Divsalar_ITA10}
S.~Abu-Surra, D.~Divsalar, and W.~E. Ryan, ``On the existence of typical
  minimum distance for protograph-based {LDPC} codes,'' in \emph{Information
  Theory and Applications Workshop (ITA)}, Jan. 2010, pp. 1 --7.

\bibitem{Liva2007}
G.~Liva and M.~Chiani, ``Protograph {LDPC} codes design based on {EXIT}
  analysis,'' in \emph{Proc. IEEE GLOBECOM}, Nov. 2007, pp. 3250--3254.

\bibitem{Hu03_PEG}
X.-Y. Hu, E.~Eleftheriou, and D.-M. Arnold, ``Regular and irregular progressive
  edge-growth {T}anner graphs,'' \emph{{IEEE} Trans. Inform. Theory}, vol.~51,
  pp. 386--398, 2003.

\bibitem{SmarandacheVontobel:IT2012}
R.~Smarandache and P.~O. Vontobel, ``Quasi-cyclic {LDPC} codes: {I}nfluence of
  proto- and {Tanner}-graph structure on minimum {Hamming} distance upper
  bounds,'' \emph{{IEEE} Trans. Inform. Theory}, vol.~58, pp. 585 -- 607, Feb.
  2012.

\end{thebibliography}
\end{document}